\documentclass{article}
\usepackage{smc2026}
\usepackage[caption=false, font=footnotesize]{subfig}
\usepackage{paralist}
\usepackage[figure,table]{hypcap}

\usepackage[english]{babel}

\usepackage{inconsolata}

\usepackage{newtxtext, newtxmath}

\usepackage{microtype}

\usepackage[noabbrev]{cleveref}

\usepackage{siunitx}
\usepackage{tikz}
\usetikzlibrary{arrows.meta}

\def\papertitle{Ambisonics: Introducing an Accessible, Distributed Approach}

\author[1]{\mbox{\firstname{Thomas}\middlename{Albert}\lastname{Rushton}}\email{thomas.rushton@inria.fr}}
\author[1]{\mbox{\firstname{Romain}\lastname{Michon}}}
\author[1]{\mbox{\firstname{Tanguy}\lastname{Risset}}}
\author[1]{\mbox{\firstname{Clemens}\lastname{Wegener}}}
\author[2]{\mbox{\firstname{Pierre}\lastname{Lecomte}}}

\affil[1]{
  \institution{Inria, INSA Lyon}
  \streetaddress{CITI, EA3720}
  \postcode{69621}
  \city{Villeurbanne}
  \country{France}
  \affiliationtype{Research}
}

\affil[2]{
    \institution{École Centrale de Lyon}
    \streetaddress{36 Av. Guy de Collongue}
    \postcode{69130}
    \city{Écully}
    \country{France}
    \affiliationtype{Research}
}

\completesetup

\title{\papertitle}
\begin{document}
	\capstartfalse
	\maketitle
	\capstarttrue

	\begin{abstract}
        This paper presents an accessible, distributed approach to higher order ambisonics using a network of low-cost microcontroller-based devices. Leveraging recent advances in embedded computing, networked audio, and precise time synchronization, the system offers an alternative to conventional centralized architectures that typically require expensive multichannel interfaces and high-performance digital signal processing hardware. The proposed architecture combines an audio plugin for soundfield design with a modular network of ambisonics processing units, which, exploiting the parallelisable nature of ambisonics decoding, are each responsible for a subset of loudspeaker outputs. Synchronization is achieved through the IEEE 1588 Precision Time Protocol, complemented by adaptive clock correction to maintain sub-microsecond alignment between devices. A third-order ambisonics implementation demonstrates the viability of the approach, with potential scalability to higher orders and larger arrays. Overall, this work contributes to the development of frugal, distributed solutions for immersive audio, expanding access to ambisonics-based systems for a wide range of applications.
	\end{abstract}

	\section{Introduction}\label{sec:introduction}

    Large-scale spatial and immersive audio systems remain exclusive commodities. The cost
    of loudspeakers aside (though various frugal and DIY solutions have been 
    proposed~\cite{michonMaking2023, weschLOTTO2025}), supporting hardware --- highly 
    multichannel audio interfaces and powerful, centralised DSP systems --- is, for many
    researchers, artists, and other interested parties, prohibitively expensive; the barriers 
    to entry to implementing wave field synthesis (WFS)~\cite{berkhoutAcoustic1993} and 
    higher order ambisonics (HOA)~\cite{zotterAmbisonics2019} systems remain stubbornly high.

    Research in recent years has, however, witnessed efforts to lower these barriers to entry.
    Exploiting advancements in embedded and distributed computing, networked audio, plus a 
    general spirit of frugality and DIY-mentality, experimental and low-cost spatial audio
    systems have been developed in attempts to provide viable alternatives to the commercial
    status quo~\cite{devonportDistribution2019, bellochPerformance2021, rushtonMicrocontrollerbased2023, michonMaking2023, rushtonNetworked2024, quiedevilleSpace2025, rushtonAll2025, weschLOTTO2025}.

    In this paper we present a distributed HOA system consisting of a digital audio 
    workstation (DAW) plugin and a synchronous network of low-cost microcontroller (MCU)
    based devices. Exploiting the parallelisable nature of the ambisonics decoding process, 
    the system is scalable and modular, with a minimum of redundancy, and provides a flexible 
    framework for spatial audio reproduction, while highlighting remaining challenges in 
    configuration and channel scalability. In principle any parallelisable spatial audio algorithm could be implemented on our system; here, in \cref{sec:implmentation}, we give the example of a
    third order HOA installation, adding, in \cref{sec:conclusion}, that 
    it is potentially scalable to at least fifth order.

    \section{Background}

    A variety of comparatively recent technological advancements have made the present work
    possible. These include the rise of embedded computing and \emph{internet of things}
    platforms, the adoption of networked audio as the standard in multichannel audio 
    transmission in local area settings, the introduction of the IEEE 1588 Precision Time
    Protocol (PTP), and much research in spatial and immersive audio. In the following
    subsections we describe the most pertinent contributing developments.

    \subsection{Frugal Spatial Audio}

    Recent years have seen various attempts to democratise and render spatial and immersive
    audio installations more accessible. These include frugal wave field synthesis arrays
    driven by a Field Programmable Gate Array 
    (FPGA)~\cite{michonMaking2023, quiedevilleSpace2025} and the authors' previous work on
    distributed sound field synthesis systems based on networked 
    MCUs~\cite{rushtonMicrocontrollerbased2023, rushtonNetworked2024, rushtonAll2025}.
    Due to the parallelisable nature of sound field synthesis algorithms, and their 
    adeptness in parallel computation, the FPGA-based approach shows promise, but it is 
    encumbered by the relative cost of FPGA development boards, plus the difficulty 
    associated with programming them. High level tools for programming FPGAs, such as the 
    SyFaLa toolchain~\cite{popoffAudio2023}, do exist, but nonetheless the complexity 
    associated with FPGA development poses at the very least a steep learning curve.

    The audio DIY and maker communities have benefited in the past decade or so from the
    emergence of a variety of \emph{embedded} platforms for audio development, such as
    MCU-based devices --- ESP32 and STM boards, plus the Daisy and Teensy --- and embedded
    Linux systems --- the Bela, Elk, and Raspberry Pi amongst the 
    latter~\cite{michonEmbedded2020}. These platforms offer a variety of audio-specific
    advantages: suites of audio-specific libraries in the case of the Daisy, Teensy and 
    Bela, programmability in the high-level audio domain specific language (DSL) 
    Faust in those and many other such cases~\cite{michonEmbedded2020}, very low latency
    in the case of the Bela platform, and all-in-all a low-cost, small form-factor way of
    getting sophisticated audio processing and control interfaces into novel digital 
    musical instruments and installations.

    \subsection{Distributed Spatial Audio}

    A fundamental accessibility issue by which spatial audio systems are afflicted is that
    of their typically centralised nature. In order to cover a large listening area with
    optimal spatial resolution, a large number of loudspeakers must be employed, which in 
    turn traditionally entails the deployment of high channel-count audio interfaces. Even
    a comparatively small-scale WFS system may demand as many as 64 output 
    channels~\cite{graniGestural2016}; a fifth-order HOA system calls for a minimum of 
    \((5+1)^2 = 36\) channels, and ideally, to maximise the system's sweet-spot listening
    area, many more. Recent years have augured in efforts to \emph{distribute} the problem
    of spatial audio, with both HOA~\cite{devonportDistribution2019} and
    WFS~\cite{bellochPerformance2021,rushtonMicrocontrollerbased2023,rushtonNetworked2024, rushtonAll2025}
    given due attention.

    Distributed audio processing gives rise to inevitable problems around synchronisation.
    No two devices configured to run at a given sampling rate will run at exactly the same
    precise rate due to the manufacturing tolerance of their internal time reference, a
    crystal oscillator. It has been demonstrated that MCU-based devices possessing the 
    appropriate functionality can be synchronised to sub-microsecond order via PTP~\cite{carotGNSSbased2019,schleusnerSubMicrosecond2024},
    and furthermore, that PTP-based time exchange can be used for media clock and 
    presentation time recovery~\cite{rushtonAll2025}.

    Despite being memory-poor, sporting only 1024 KB RAM~\cite{rushtonAll2025}, a Teensy
    4.1 MCU development board equipped with audio and ethernet add-ons costs less than 
    € 50~\cite{rushtonAll2025}, and memory can be extended by up to \qty{16}{\mega\byte} 
    by way of inexpensive surface-mount PSRAM 
    chips.\footnote{\url{https://www.pjrc.com/store/psram.html}} Furthermore, its NXP
    IMXRT1062 MCU~\cite{nxpsemiconductorsIMX2021} features an ARM core processor clocked 
    at \qty{600}{\MHz}, more powerful than many similar MCU-based 
    platforms~\cite{rushtonNetworked2024}. Critically, for purposes of synchronisation in 
    a distributed setting, the IMXRT1062 also features a PTP timer, hardware timestamping 
    functionality, and a dedicated audio subsystem with extremely high-resolution 
    phase-locked loop (PLL) numerator and divider registers --- in effect it is possible 
    to enact nanosecond-order adjustments to the sampling rate of the device during 
    runtime to ensure synchronous reproduction across networked modules; indeed, although
    commercial multichannel audio interfaces exist at a pricepoint of € 150 for ten line 
    outputs (e.g. Behringer UMC1820), if using multiple such interfaces for high 
     channel-count applications there is no way to synchronise the devices. Furthermore, 
    being programmable in Faust, the Teensy can benefit from the existence of the 
    \emph{AmbiTools} suite of Faust-based utilities for 
    ambisonics~\cite{lecomteAmbitools2018}.

    \subsection{Ambisonics}

    Ambisonics is a sound field synthesis technique based on the principle of decomposing a
    sound field onto a truncated basis of spherical harmonics~\cite{zotterAmbisonics2019}; the
    \emph{order} \(M\) of the basis truncation gives the total number of spherical harmonics
    used to decompose the sound field: \((M + 1)^{2}\). Using a limited number of functions 
    for the decomposition limits the spatial zone where the sound field can be accurately
    represented and reconstructed. Its size is such that \(\lceil kr \rceil = M\), where 
    \(r\) is the radius of the zone of accurate reconstruction (the \emph{``sweet spot''}) and
    \(k = 2 \pi f / c\) is the wave number, \(f\) is frequency, and \(c\) the speed of sound.
    
    First order ambisonics, for which \(M = 1\), uses four signals, and was first theorized 
    and developed by Gerzon~\cite{zotterAmbisonics2019}. HOA is implemented when \(M > 1\). 
    For sound field reconstruction the number of loudspeakers required is 
    \(N \ge (M + 1)^2\). Estimation of the ambisonic components of a sound pressure field is 
    possible via recording with spherical microphone arrays~\cite{moreau3D2006};
    alternatively, monophonic sound sources can be spatialised via \emph{encoding} to 
    ambisonic format~\cite{lecomteAmbitools2018}, which, by convention, uses \emph{ambisonics 
    channel numbering} (ACN) to designate channels. 
    
    ACN channels are indexed by the degree \(m\) and order \(n\) of the Legendre polynomials by which they are described:~\cite{lecomteAmbitools2018}

    \begin{equation}
        a = m^{2} + m + n\;.
    \end{equation}
    For a single monophonic source \(x(t)\) at azimuth \(\theta\) and elevation \(\varphi\), the encoded signal \(b_{a}\) for a given ACN-format channel is

    \begin{equation}
        b_a(t) = x(t) \cdot Y_{m, n}(\theta, \varphi)\;,
    \end{equation}
    where \(Y_{m, n}\) is the real valued spherical harmonic~\cite{lecomteAmbitools2018}

    \begin{equation}
        Y_{m, n}(\theta, \varphi) = N_{m, n}P_{m, |n|}\sin(\varphi) \begin{cases}
            \cos(|n|\theta), \; n \ge 0 \\ \sin(|n|\theta), \; n < 0
        \end{cases}.
    \end{equation}
    Here \(N_{m, n}\) is a normalisation factor, and \(P_{m, |n|}\) are the associated 
    Legendre polynomials. Ambisonics is not without its terminological
    ambiguities~\cite{nachbarAmbiX2011}: it is the \emph{degree} of the polynomial that 
    corresponds with the ambisonics \emph{order}. 
    
    Considering the case of a third order ambisonics system with the bare-minimum 
    \((3 + 1)^{2} = 16\) loudspeakers at positions \(\theta_{i}, \varphi_{i}\) the 
    spherical harmonics make up the \(16 \times 16\) mode matrix 
    \(\mathbf{Y}_{m, n}(\theta, \varphi)\). In this case, the decoding matrix \(\mathbf{D}\) 
    is simply the inverse, \(\mathbf{D} = \mathbf{Y}^{-1}\), the rows of which are the 
    decoding coefficients for each loudspeaker; for a greater number of loudspeakers it would 
    be necessary to take the pseudoinverse, 
    \(\mathbf{D} = (\mathbf{Y}^{T}\mathbf{Y})^{-1}\mathbf{Y}^{T}\).

    The time-domain output signal for the \(i^{\text{th}}\) loudspeaker is found as the 
    dot product of the \(i^{\text{th}}\) row of \(\mathbf{D}\) and the \(16 \times 1\)
    vector of ACN-format channels:

    \begin{equation}\label{eq:ithspeakeroutput}
        s_{i}(t) = \mathbf{D}_{i} \cdot \mathbf{b}(t)\;,
    \end{equation}
    and it is clear, then, that the ambisonics decoder is parallelisable --- a DSP module
    producing output for speakers \(i\) and \(j\) need only be aware of the coefficients 
    for the corresponding \(i^{\text{th}}\) and \(j^{\text{th}}\) rows of the decoding
    matrix.

    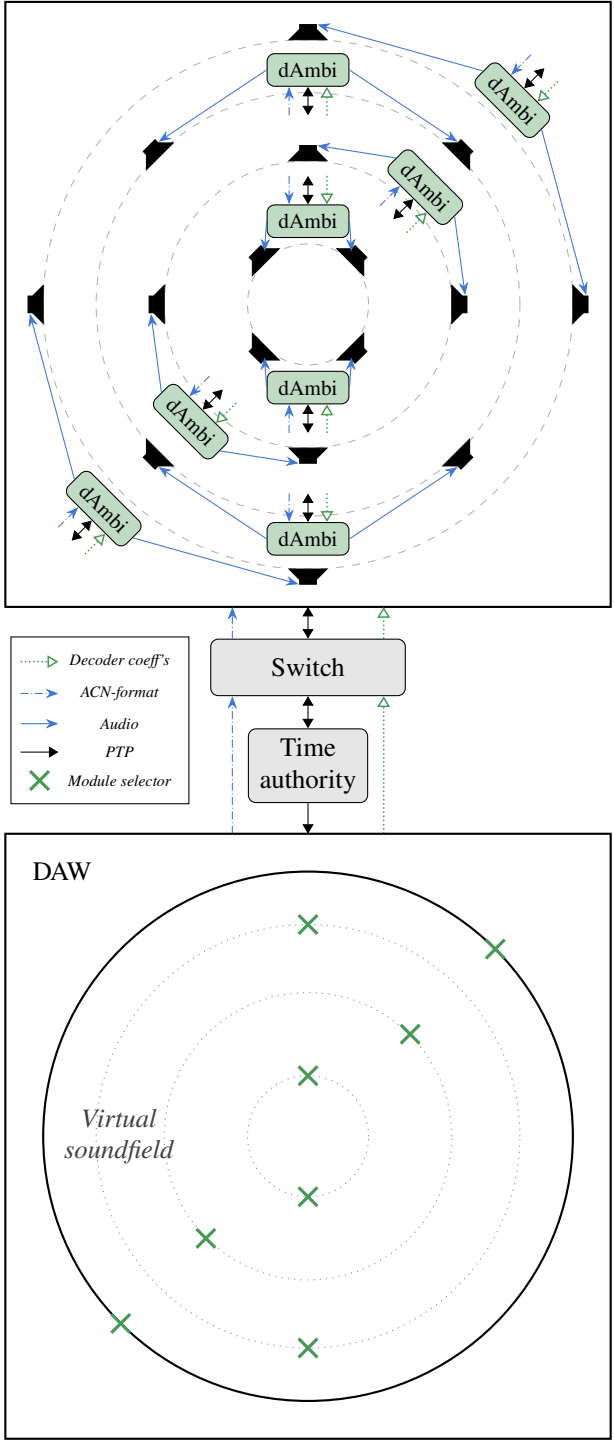
\begin{figure}[ht!]
        \definecolor{sourceColour}{HTML}{4077D6}
        \definecolor{ctrlColour}{HTML}{479955}
        \centering
        \begin{tikzpicture}[
            device/.style={draw, fill=black!10, rounded corners=1mm, inner xsep=1.5mm},
            ctrl/.style={-{Triangle[open]}, densely dotted, ctrlColour, line width=.2mm},
            bformat/.style={-{Stealth}, sourceColour, densely dashdotted},
            audio/.style={-{Stealth}, sourceColour},
            ptp/.style={{Triangle}-{Triangle}},
            legend/.style={font=\tiny\itshape},
            module/.style={device, fill=ctrlColour!35, font={\footnotesize}, transform shape},
            selector/.pic = {
                \draw[rotate=45, very thick, ctrlColour] (-#1,0) -- (#1,0);
                \draw[rotate=45, very thick, ctrlColour] (0,-#1) -- (0, #1);
            }
        ]
            \begin{scope}[
                yshift=11cm,
                speaker/.pic = {
                    \draw [fill=black] (-.11, 0) rectangle (.11, .06);
                    \draw [fill=black] (-.11, .06) -- (-.25, .2) -- (.25, .2) -- (.11, .06);
                    \node (-body) at (0, 0) {};   
                },
                moduleN/.pic = {
                    \node [module] (-body) {dAmbi};
                    \draw [ptp] (0, .6) -- (-body.north);
                    \draw [bformat] (-.25, .6) -- ([xshift=-.25cm]-body.north);
                    \draw [ctrl] (.25, .6) -- ([xshift=.25cm]-body.north);
                },
                moduleS/.pic = {
                    \node [module] (-body) {dAmbi};
                    \draw [ptp] (0, -.6) -- (-body.south);
                    \draw [bformat] (-.25, -.6) -- ([xshift=-.25cm]-body.south);
                    \draw [ctrl] (.25, -.6) -- ([xshift=.25cm]-body.south);
                }
            ]
                \draw [thick] (-4, -4) rectangle (4, 4);
                \draw [dashed, black!30] (0, 0) circle (3.5);
                \draw [dashed, black!30] (0, 0) circle (2.8);
                \draw [dashed, black!30] (0, 0) circle (1.9);
                \draw [dashed, black!30] (0, 0) circle (.8);
                \pic [rotate=180] (speaker0) at (0, 3.7) {speaker};
                \pic [rotate=90] (speaker1) at (3.7, 0) {speaker};
                \pic (speaker2) at (0, -3.7) {speaker};
                \pic [rotate=-90] (speaker3) at (-3.7, 0) {speaker};
                \pic [rotate=225] (speaker4) at (-2.1, 2.1) {speaker};
                \pic [rotate=135] (speaker5) at (2.1, 2.1) {speaker};
                \pic [rotate=45] (speaker6) at (2.1, -2.1) {speaker};
                \pic [rotate=-45] (speaker7) at (-2.1, -2.1) {speaker};
                \pic [rotate=180] (speaker8) at (0, 2.1) {speaker};
                \pic [rotate=90] (speaker9) at (2.1, 0) {speaker};
                \pic (speaker10) at (0, -2.1) {speaker};
                \pic [rotate=-90] (speaker11) at (-2.1, 0) {speaker};
                \pic [rotate=225] (speaker12) at (-.7, .7) {speaker};
                \pic [rotate=135] (speaker13) at (.7, .7) {speaker};
                \pic [rotate=45] (speaker14) at (.7, -.7) {speaker};
                \pic [rotate=-45] (speaker15) at (-.7, -.7) {speaker};
                \pic [rotate=-45] (module0) at (2.7, 2.7) {moduleN};
                \pic [rotate=-45] (module1) at (-2.7, -2.7) {moduleS};
                \pic (module2) at (0, 3.1) {moduleS};
                \pic (module3) at (0, -3.1) {moduleN};
                \pic (module4) [rotate=-45] at (1.55, 1.55) {moduleS};
                \pic (module5) [rotate=-45] at (-1.55, -1.55) {moduleN};
                \pic (module6) at (0, 1.1) {moduleN};
                \pic (module7) at (0, -1.1) {moduleS};
                \draw [audio] (module0-body.west) -- (speaker0-body.east);
                \draw [audio] (module0-body.east) -- ([yshift=.125cm, xshift=.1cm]speaker1-body.west);
                \draw [audio] (module1-body.east) -- (speaker2-body.west);
                \draw [audio] (module1-body.west) -- ([yshift=-.125cm, xshift=-.1cm]speaker3-body.east);
                \draw [audio] (module2-body.west) -- ([yshift=.05cm]speaker4-body.east);
                \draw [audio] (module2-body.east) -- ([yshift=.05cm]speaker5-body.west);
                \draw [audio] (module3-body.east) -- ([yshift=-.05cm]speaker6-body.west);
                \draw [audio] (module3-body.west) -- ([yshift=-.05cm]speaker7-body.east);
                \draw [audio] (module4-body.west) -- (speaker8-body.east);
                \draw [audio] (module4-body.east) -- ([yshift=.125cm, xshift=.1cm]speaker9-body.west);
                \draw [audio] (module5-body.east) -- (speaker10-body.west);
                \draw [audio] (module5-body.west) -- ([yshift=-.125cm, xshift=-.1cm]speaker11-body.east);
                \draw [audio] (module6-body.west) -- (speaker12-body.east);
                \draw [audio] (module6-body.east) -- (speaker13-body.west);
                \draw [audio] (module7-body.east) -- (speaker14-body.west);
                \draw [audio] (module7-body.west) -- (speaker15-body.east);
            \end{scope}
            \begin{scope}[yshift=5.5cm]
                \node[device, inner xsep=8mm, inner ysep=2.5mm](switch) at (0, .7) {Switch};
                \node[device, align=center](authority) at (0, -.6) {Time\\authority};
                \draw [{Triangle}-] (0, -1.5) -- (authority.south);
                \draw [ptp] (authority.north) -- (switch.south);
                \draw [ptp] (switch.north) -- (0, 1.5);
                \draw [bformat] (-1, -1.5) -- ([xshift=-1cm] switch.south);
                \draw [bformat] ([xshift=-1cm] switch.north) -- (-1, 1.5);
                \draw [ctrl] (1, -1.5) -- ([xshift=1cm] switch.south);
                \draw [-{Triangle[open]}, ctrl] ([xshift=1cm] switch.north) -- (1, 1.5);
                \matrix [draw] at (-2.75, 0) {
                    \draw [ctrl] (0,0) -- (.5,0); & \node [legend] {Decoder coeff's}; \\
                    \draw [bformat] (0,0) -- (.5,0); & \node [legend] {ACN-format}; \\
                    \draw [audio] (0,0) -- (.5,0); & \node [legend] {Audio}; \\
                    \draw [-{Triangle}] (0,0) -- (.5,0); & \node [legend] {PTP}; \\
                    \pic at (.25, 0) {selector=5pt}; & \node [legend] {Module selector}; \\
                };
            \end{scope}
            \begin{scope}
                \draw [thick] (-4, -4) rectangle (4, 4);
                \node [] at (-3.25, 3.5) {DAW};
                \draw [thick] (0, 0) circle (3.5);
                \node [black!75, align=center, font={\itshape}] at (-2.5, 0) {Virtual\\soundfield};
                \draw [dotted, black!50] (0, 0) circle (2.8);
                \draw [dotted, black!50] (0, 0) circle (1.9);
                \draw [dotted, black!50] (0, 0) circle (.8);
                \pic at (2.475, 2.475) {selector=5pt};
                \pic at (-2.475, -2.475) {selector=5pt};
                \pic at (0, 2.8) {selector=5pt};
                \pic at (0, -2.8) {selector=5pt};
                \pic at (1.35, 1.35) {selector=5pt};
                \pic at (-1.35, -1.35) {selector=5pt};
                \pic at (0, .8) {selector=5pt};
                \pic at (0, -.8) {selector=5pt};
            \end{scope}
        \end{tikzpicture}
        \caption{Example third order distributed ambisonics setup. A DAW plugin describes a 
        virtual soundfield and permits the selection of distributed ambisonics 
        (\textit{dAmbi}) modules. Decoding coefficients and ACN-format encoded audio are 
        transmitted via the network switch. The time authority provides a temporal frame of
        reference for the system as a whole.}
        \label{fig:distributedambi}
    \end{figure}

    \section{Implementation}\label{sec:implmentation}

    Our system consists of a DAW plugin featuring an interface for describing a virtual sound 
    field, coupled with a network of distributed Teensy 4.1 ambisonics modules. A general-
    purpose, consumer-grade laptop computer runs the DAW which hosts the plugin, and this 
    machine is connected to a network switch, to which our Teensy-based ambisonics modules 
    are connected in turn. An additional Teensy serves as a time authority for the network 
    and the general purpose machine. \Cref{fig:distributedambi} illustrates an example third 
    order HOA installation based on our system; this system is described in further detail 
    below.

    \subsection{The Time Authority}

    Fundamental to our system is the notion of an \emph{authoritative source of time}. To
    this end, we use one Teensy as a dual-purpose USB audio device and PTP clock authority. 
    In serving this dual purpose, the time authority ensures that the computer to which it is 
    connected via USB, and the network of MCU-based ambisonics modules, for which it serves
    as the PTP authority, remain in sync. The computer, time authority, and ambisonics modules
    are all connected to a MikroTik CRS326-24G-2S+IN network 
    switch,\footnote{\url{https://mikrotik.com/product/crs326_24g_2s_in}} which acts as a PTP
    boundary clock, i.e. the switch synchronises to the time authority and the ambisonics 
    modules synchronise to the switch.
    
    \subsection{The Audio Plugin}

    The plugin was created using the CLAP\footnote{\emph{Clever Audio Plug-in}, 
    \url{https://cleveraudio.org/}} format and is hosted by a DAW running on a 
    general-purpose laptop computer. Due to its support for the CLAP format (which provides 
    greater flexibility with regard to channel layout than VST3), our preferred DAW has been
    Reaper,\footnote{\url{https://reaper.fm}} running on either Linux or Mac OS.
    
    Present in the plugin interface are elements for selecting modules, each of which is 
    responsible for producing output signals for two loudspeakers. Modules are selectable via 
    their IP addresses; once assigned a position, the plugin sends the appropriate module a 
    user datagram protocol (UDP) packet containing its decoding coefficients, i.e. the 
    \(i^{\text{th}}\) and \(j^{\text{th}}\) rows of the decoding matrix \(\mathbf{D}\) (see
    \cref{eq:ithspeakeroutput}).

    The plugin sends the ACN-format encoded ambisonics scene over UDP multicast on 
    address \texttt{224.4.224.4} --- taken from the second \emph{ad-hoc} block of the IPv4
    Multicast Address Assignment Guidelines~\cite{meyerIANA2010} from the Internet 
    Assigned Numbers Authority (IANA) --- and port 49152 --- the first port in IANA's dynamic 
    port range~\cite{cottonInternet2011}. Packets contain 16 frames of 16-bit ACN-format 
    and contain a header with a 64-bit timestamp, for synchronous reproduction, plus 8-bit 
    channel and 16-bit frame count.\footnote{We use the term \emph{frame} to refer to one 
    sample's worth of \(N\) audio channels.} The packet header is thus eleven bytes in 
    size; 16 frames of 16 channels of 16-bit ACN-format brings the overall packet size to 
    523 bytes, well within the standard 1500-byte UDP \emph{Maximum Transmissible
    Unit} (MTU).

    \subsection{Ambisonics Modules}

    \begin{figure}
        \centering
        \includegraphics[width=\linewidth]{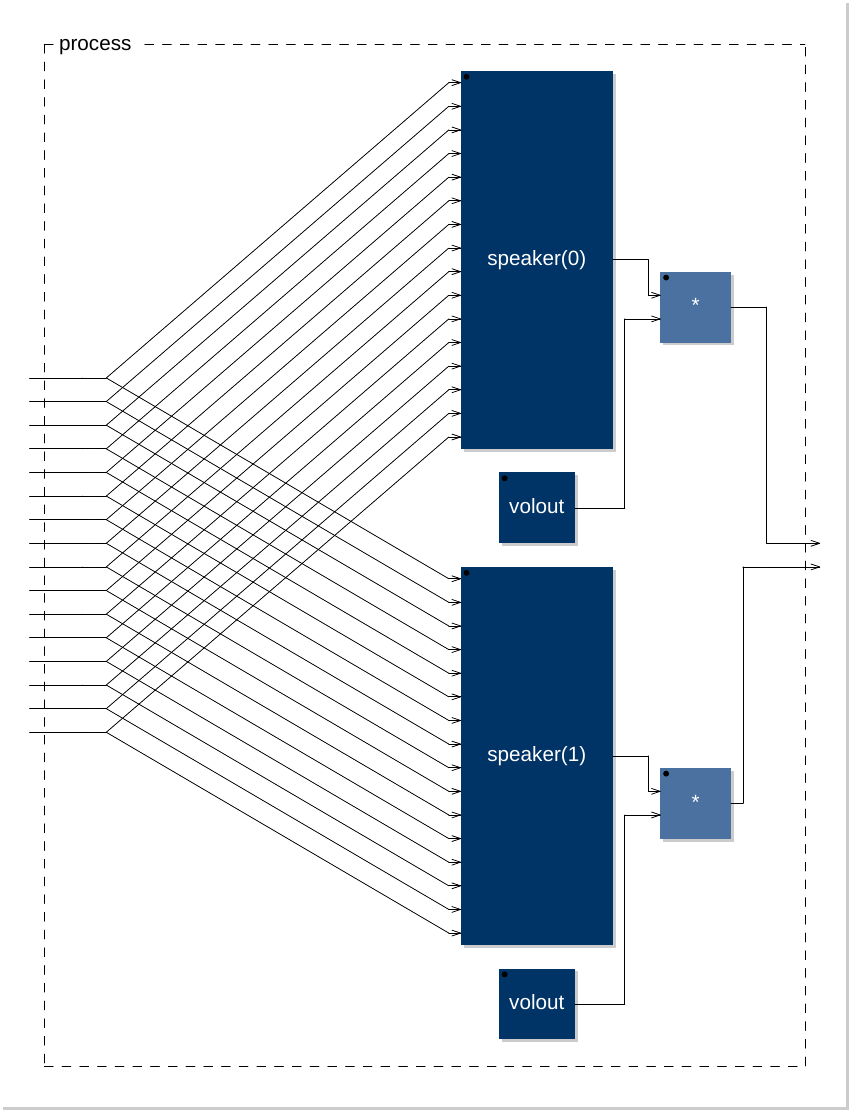}
        \caption{Faust block diagram of the distributed HOA decoder running on one of the hardware modules.}
        \label{fig:faust-algo}
    \end{figure}

    Connected to the switch are our eight MCU-based ambisonics modules. 
    Modules are 
    synchronised according to the method described in our previous 
    work~\cite{rushtonAll2025}, with a further measure in place to ensure 
    sub-microsecond drift: in addition to using PTP-derived media clock recovery, modules
    check the offset of their audio subsystem from PTP time once per second and make a
    commensurate sampling rate adjustment. Using this hybrid adaptive clock-correction
    method, presentation remains synchronous amongst the network of modules to within around 
    \qty{200}{\nano\second} at all times.
    The generated soundfield has been subjected to preliminary tests to ascertain its plausibility; however, no qualitative measurements or empirical assessments have been conducted at the time of writing.
    
    An adapted version of the \emph{AmbiTools} Faust algorithm is used to program the 
    modules, taking advantage of the aforementioned parallelisation of the 
    ambisonics decoder. See \cref{fig:faust-algo} for a diagrammatic representation of the 
    distributed HOA decoder.

	\section{Conclusions and Outlook}\label{sec:conclusion}

    In this paper we have demonstrated, taking advantage of the parallelisable nature of 
    the decoding step of ambisonic reproduction, a distributed ambisonics system based on 
    a low-cost MCU-based embedded computing platform. Our aim is that this should 
    represent an important step toward a more accessible future for ambisonics-based 
    immersive sound installations.

    Though functional, scalable and modular, our system requires a comparatively laborious
    configuration step, that of assigning modules to their positions in the speaker array;
    ideally an automatic calibration mode, such as demonstrated by Wesch et 
    al.~\cite{weschLOTTO2025} should be implemented to render this less of a pain-point.
    Moreover, the system presented uses the Teensy's standard audio add-on, which provides
    two channels of output; via time-division multiplexing (TDM), it should be possible to 
    produce as many as 32 output channels per module, significantly simplifying the
    overall design providing a more favourable cost-per-output-channel.

    In principle, without violating the MTU, we could support fifth order ambisonics,
    transmitting 36 ACN-format channels for a packet size of 1163 bytes. For optimal rather
    than bare-minimum reproduction, however, needing something on the order of fifty 
    loudspeakers for a hemispherical configuration, at two speakers per module, and given 
    the 24 ethernet ports of the CRS326-24G-2S+IN switch, this would entail daisy-chaining 
    switches; since the switch (chosen for its low cost, high port-count, and 
    PTP-compliance), is only capable of acting as a boundary clock, modules connected to the 
    second switch would be subject to a delay of perhaps a few hundred nanoseconds. Again, 
    channels-per-module improvements via TDM would sidestep this issue. Nonetheless, that, 
    and other technical and logistical issues remain to be explored in future work.

    The work presented here forms part of a broader project engaged with frugal, distributed
    approaches to spatial audio and virtual acoustics. Full details of the hardware,
    supporting software, and network protocol, including synchronisation, device discovery,
    and audio and control data transmission, will be made public and freely available as the
    project develops further.
	
	\bibliography{smc2026bib}
	
\end{document}